# A first study on iron complexes from blood and organ samples of thalassaemic and normal lab-mice via Mössbauer Spectroscopy

George Charitou[1,a], Vlassis Petousis[1,b], Charalambos Tsertos[1], Yannis Parpottas[2], Marina Kleanthous[3], Marios Phylactides[3], Soteroula Christou[4]

[1] *Department of Physics, University of Cyprus, Nicosia 1678, Cyprus*

[2] *Frederick University, Nicosia 1036, Cyprus*

[3] *Department of Molecular Genetics Thalassaemia, Cyprus Institute of Neurology and Genetics, Nicosia 1683, Cyprus*

[4] *Thalassaemia Center, Archbishop Makarios III Hospital, 1474 Nicosia, Cyprus*

(Revised version: 16/05/2017)

**Abstract** In this paper, blood and tissues samples from one normal and one thalassaemic lab-mice were studied using $^{57}$Fe Mössbauer spectroscopy at 78K for the first time. In contrast to human patients, these laboratory mice did not receive any medical treatment, thus the iron components present in the samples are not altered. The measured Mössbauer spectra of the blood, liver and spleen samples of the thalassaemic mouse were found to be different in the shape and iron content as compared to the corresponding spectra of the normal mouse. This result demonstrates the further exploitation of the thalassaemic mouse model to study thalassaemia in more details by means of Mössbauer spectroscopy.

**Introduction**

Beta-thalassaemia (Mediterranean anaemia) is a form of inherited autosomal recessive blood disorder characterized by reduced or absent β-globin chain synthesis. This anomaly prevents the production of normal levels of adult haemoglobin and also leads to the generation of unstable α-globin chain aggregates, which in turn limit the effective production of Red Blood Cells (RBCs) leading to mild or severe anaemia.

Haemoglobin is a highly specialized metalloprotein which is responsible to transport oxygen from the lungs to the tissues. It is found in abundance in the RBCs. Each RBC contains about 300 million molecules of haemoglobin and it is formed by two pairs of identical subunits, the globin chains (Cappellini, Cohen, et al. 2014). Each chain consists of a protein part, and a haeme group which consists of an organic compound called protoporphyrin and a central iron atom (Berg et al. 2002).

In healthy humans, the iron that is not used by the body is attached to transferrin, which is an iron binding glycoprotein found in the serum (Crichton and Charloteaux-Wauters 1987) and it circulates within the blood. It is also stored as ferritin and haemosiderin in the liver, spleen and bone marrow (Yutaka et al. 2008). Ferritin is a spherical molecule with an 8 nm central cavity which holds 2000 – 4500 iron (III) oxy-hydroxide atoms (Chasteen and Harrison 1999). It is an iron storage protein which keeps the iron in a soluble and non-toxic form while haemosiderin is an insoluble form of tissue storage iron. As the iron level in the body increases beyond normal level, the number of ferritin and haemosiderin molecules are increased, as well as the iron core size with the number of iron ions in the core.

In thalassaemia and haemochromatosis, iron can exist in forms not bound to transferrin (NTBI: Non Transferrin Bound Iron) or other traditional binding proteins like haem, ferritin, haemosiderin (Patel and Ramavataram 2012), due to an increased transferrin saturation (Yutaka et al. 2008; Eleftheriou 2003). NTBI can catalyse the Fenton and Haber-Weiss chemical reactions (Patel and Ramavataram 2012; Eleftheriou 2003; Evans et al. 2008), and therefore it could produce harmful hydroxyl radicals which are potentially toxic to cells (Evans et al 2008; Brissot, Ropert et al. 2012) and can cause extensive damage to the body tissues (Eleftheriou 2003).

Patients with severe anaemia are heavily dependent on frequent blood transfusions to ensure oxygen transport, which leads to iron overload (Cappellini, Cohen, et al. 2014). Further, the patients in mild anaemia may also suffer from iron overload due to increased gastrointestinal absorption of iron from the diet (Berg et al. 2002) since there is no active mechanism to excrete iron from the body (Yutaka et al. 2008).

---

[a] Part of his PhD Thesis, email: haritou.georgios@ucy.ac.cy

[b] petousis@ucy.ac.cy

Even with the most modern treatments, patients face life threatening complications, bone deformities, poor growth, hepatosplenomegaly and cardiovascular illness which reduce their life expectancy. Chelating agents, such as desferrioxamine and deferiprone, are used to remove the excess iron from the body. Accurate, preferably non-invasive, measurements of iron complexes and iron stores in the body are crucial for the evaluation and management of chelation therapy.

$^{57}$Fe Mössbauer spectroscopy (Vertes, Korecz and Burger 1979; Frauenfelder 1963; Goldanskii and Herber 1968; Greenwood and Gibb 1971; Gutlich, Link and Trautwein 1978; Gutlich, Bill and Trautwein 2011)(MS) is capable of characterising these iron complexes and in particular to provide information such as the iron electronic structure, magnetic structure, hyperfine interactions, valence/spin state, local microenvironment / iron stereochemistry and symmetry of environment, iron bonding, number of resonant nuclei, dynamics. This in contrast with the chemical or clinical markers, which are mainly dedicated to detecting specific iron complexes. Accurate characterization of iron complexes in the body may lead to the improvement of iron chelating agents.

Previous Mössbauer studies characterized iron complexes in blood and organs of humans and of some animals. In particular, comparing RBCs from healthy people and patients with thalassaemia, a new component was identified in the patient spectra which was considered to be ferritin-like iron (Abreu, Sanchis et al. 1989; Xuanhui, Nanming et al. 1988; Jiang, Ma et al. 1994). Other studies considered the differences in the haeme iron electronic structure in both oxy- and deoxyhemoglobins, in normal adult, fetal and leukemic RBCs. They presented the α- and β-subunits of oxyhaemoglobin (Oshtrakh 1998; Oshtrakh, Berkovsky et al. 2011) in normal human RBC samples. When heart, spleen and liver tissues from healthy individuals and patients with thalassaemia were compared using $^{57}$Fe MS, an increased quantity of ferritin and haemosiderin was found in the thalassaemic samples (Kaufman at al. 1980; Bell et al. 1984; Rimbert et al. 1985; Pierre et al. 1998; Chua-anusorn et al. 1994).

Small variations of $^{57}$Fe MS parameters were observed in concentrated normal RBC samples between the oxyhaemoglobins (α- and β-subunits) of a human, a rabbit and a pig, which are related to well-known small structural variations in the haeme iron stereochemistry (Oshtrakh, Berkovsky et al. 2011; Oshtrakh, Kumar et al. 2011). Further, ferritin and haemosiderin were identified in dietary-iron-loaded and parenteral-loaded liver and spleen samples from rats (Chua-anusorn et al. 1999). It was also found that haemosiderin levels increased with the age of the rat and hence the duration of iron overload (Chua-anusorn et al. 1999). Also, the effect of the iron chelating drug desferrioxamine (Desferal) was studied revealing that it could remove a major part of iron aggregated from the iron overload in rat myocardial cells (Bauminger et al. 1987).

Due to the severe complications of thalassaemia, patients receive blood transfusions and iron-chelators from young age. These alter the various iron forms and their concentration in the body. Therefore, characterization of the effect of thalassaemia alone on such iron complexes is not possible using human samples. Thalassaemic mice have been used in various fields of research since they develop iron deposits very early (Yang, Kirby et al. 1995) and therefore they can be suitable candidates to study iron accumulations in the organs using $^{57}$Fe MS. A direct comparison between normal and thalassaemic mice may identify iron complexes arising due to thalassaemia and may reveal ways of iron accumulation or overload in blood and organs. This could lead to the design of more appropriate binding agents and to the production of more efficient iron chelators, reducing the effects of iron overload in the patients.

Thus, in this paper, Mössbauer spectra of blood, liver and spleen samples from one normal and one thalassaemic lab-mice were acquired and compared in order to provide evidence whether this thalassaemic mouse model can be used for characterizing iron complexes associated with thalassaemia.

**Materials and methods**

**Experimental Setup**

A complete Mössbauer spectroscopy equipment provided by WissEl (WisseEl et al. 2011) was utilized for our measurements and the spectra were recorded in transmission geometry. A 50 mCi $^{57}$Co/Rh source was driven with a constant acceleration using a triangular mode which was generated by a Digital Function Generator for the velocity of ±4 mm/sec.

The 14.41 keV γ-rays of the $^{57}$Fe nucleus were detected using Xenon (Xe)-filled proportional counter (LND 4546). The desired sample was placed inside a dedicated holder and then inserted into the cryostat (Ice$^{Bath}$ Ice Oxford). The cryostat was filled with liquid nitrogen (LN) and operated at a vacuum of $7\times10^{-7}$ mbar. A thermocouple which was attached on the sample holder was used to measure the sample temperature. The temperature was held stable, within ±0.5K, using a controller and a heater which was attached to the cryostat's heat exchanger.

The Mössbauer spectrum of each biological sample was acquired for about 15 days to obtain a statistics of about $1\times10^{6}$ counts per channel. 1024 channels were used to register the spectra (forward and backward movements) measured with a triangular velocity signal. For the analysis, the spectra were folded into 512 channels and the peaks were fitted using Lorentzian distributions. The folding and the analysis were performed using the IGOR-WinNORMOS software package of WissEl.

**Calibration**

The velocity was calibrated using the four inner peaks of a 99.85% in purity α-Fe foil (Goodfellow Cambridge Limited) with a thickness of 8 μm, at room temperature (RT, 298K). The isomer shift (δ) value of the α-Fe foil relative to the $^{57}$Co/Rh source at RT was found to be -0.110 mm/s, consistent with the source manufactured specifications (chemical shift of the source relative to α-Fe is 0.108 mm/s) and the corresponding literature (Vértes, Nagy et al. 2011).
MS is very sensitive to vibrations that might arise from the vacuum pumps and/or from the evaporation of LN inside the cryostat, so it is crucial to ensure that the resonant absorption line widths remain stable and within acceptable values (<0.30 mm/s) at the desire working conditions. For this reason, the stability of the system against vibrations has been carefully investigated by conducting measurements using a 10 mg/cm$^2$ natural abundance Fe powder sample (<10 μm, ≥99%, Sigma-Aldrich), before and after each biological-sample measurement. From such measurements, the line widths were found to lie in the range of 0.25-0.28 mm/s.
Table 1 shows the measured peak position (X) and the line width (Γ) for the α-Fe foil (RT) and from the iron powder sample (RT and 78K), as well as the corresponding reference values (Maddock 1998) (RT) for the expected peak positions. As can be seen, the measured positions of the four absorption Fe peaks are reproduced with an accuracy of a few ~10$^{-3}$, while their widths are measured with a resolution (FWHM) of about 0.25-0.28 mm/s. Figure 1 shows the measured spectra, fitted simultaneously with four Lorentzian distributions, from the Fe-powder sample: (a) at RT and (b) at 78K, together with the residuals of the fits. The overall instrumental error derived from the Fe-powder sample measurements was found to be <0.03 mm/s. In the following, the isomer shift δ values and peak positions are given relative to the α-Fe foil at RT.
Last, background measurements were carried out in addition using an empty sample holder, with a statistics of 6x10$^6$ counts per channel before folding. They have shown that there is no significant MS absorption (<0.1%) due to iron traces of the setup environment such as the sample holder and the detector window.

**Sample Holders**

Custom-made holders for these bio-samples were fabricated though the thermo-forming method using copper molds and polypropylene (PP) sheets of 0.75 mm in thickness. Cylindrical holders of different size in diameter (10 mm and 7 mm) have been used for the blood, liver and spleen samples, to better fit the corresponding sample volume within the holder. The volumes of the holders were 1, 0.5 and 0.5 ml for the blood, liver and spleen, respectively. A 2-mm-thick collimator made of copper was attached to each holder.
The absence of atoms such as oxygen, nitrogen and chlorine in the polypropylene molecule ($C_3H_6$)$_n$ compared to other thermoplastics results in high transparency at low-energy of X- and γ-rays. Table 2 shows the measured Full-Width at Half-Maximum (FWHM), the net count rate (Rate), and the percentage transmission (T) without a holder, with an existing standard Plexiglas (PMMA) holder, and our custom-made polypropylene (PP) holder at $^{57}$Co/Rh energies of 6.36 keV and 14.41 keV, respectively. The corresponding spectra are shown in figure 2. The measured transmissions were obtained using an Amptek XR-100CR Si-PIN X-Ray detector due to its high energy resolution. As it can be seen in Table 2, the X-ray at 6.36 keV cannot be transmitted through the PMMA holder. Even more importantly, the transmission through our PP holder at the energy of interest of 14.41 keV is significantly enhanced (96%) compared to that of the standard PMMA holder (56%).

**Mouse model and Sample Preparation**

Blood samples and wet tissues of liver and spleen samples from one normal (C57BL/6J) and one thalassaemic mice were investigated in this study. The thalassaemic mouse model (Yang, Kirby et al. 1995) used has deleted both the *b1* and *b2* adult mouse globin genes. Mice heterozygous for the deletion (*Hbb$^{th-3}$/Hbb$^{wt}$*) were used for the experiment as homozygous mice die perinatally. Heterozygous *Hbb$^{th-3}$/Hbb$^{wt}$* mice appear normal, but show haematologic indices characteristic of severe thalassaemia, exhibit tissue and organ damage typical of the disease and show spontaneous iron overload in the spleen, liver, and kidneys.
The two mice were age- and sex-matched and they were raised in similar conditions with the same diet provided to them. Blood samples were collected in plain Eppendorf tubes by retro-orbital bleeding of anaesthetised animals. Heparin was added in order to prevent coagulation. Following euthanasia of the animals, the liver and the spleen were isolated and washed with a phosphate buffered saline solution in order to remove excess blood traces. All samples were then placed in the custom-made holders at the Department of Physics of the University of Cyprus and stored in a Dewar at LN temperature.

**Results and Discussion**

**Blood**

Figures 3a and 3b show the MS spectra at 78K of the blood samples for the normal and the thalassaemic mice, respectively. The spectrum of the normal sample (Fig. 3a) with a signal-to-noise ratio (S/N) of 23 was fitted with two quadrupole doublets, representing the α- and β-chains of oxy-haemoglobin. Herewith, a normalized $\chi_\nu^2$ value ($\chi_\nu^2 = 0.93$) was obtained for the quality of the fit. The parameters extracted of these two quadrupole doublets were then used to fit the spectrum of the thalassaemic sample (Fig. 3b), allowing only the doublets area to vary. In this case, a higher $\chi_\nu^2$ value ($\chi_\nu^2 = 1.22$) was calculated, which along with the corresponding fit residuals shown, indicate a non-satisfactory fit. This could be due to more iron-containing complexes, present in this sample. Because of the low (S/N) ratio (~10) of this spectrum (Fig. 3b), a further more detailed analysis would not be useful.

A decreased absorption effect is observed in the spectrum (Fig. 3b) of the thalassaemic sample due to reduced haemoglobin level. This indicates extreme anaemia, which is expected from a thalassaemia major patient without any treatment (Eleftheriou 2003).

Table 3 shows the Mössbauer parameters for the fit in figure 3a. The values of these parameters are similar to those reported in other studies which measured normal RBC samples of a human, a rabbit and a pig and normal human RBCs (Oshtrakh 1998). Also, the relative area of both the α- and β-subunits is comparable in amount, which is in agreement with the equal distribution of the iron nuclei in each type of subunits in haemoglobin tetramer (Oshtrakh 1998).

A ferritin component was identified in a previous MS study, at about 80K, from RBC samples of patients with β-thalassaemia intermedia (Bauminger, Cohen et al. 1979), where the ratio of ferritin to the haemoglobin iron varied from 3-50% (Bauminger, Cohen et al. 1979). Due to the fact that thalassaemia patients undergo long-term medical treatment, a ferritin component could not be observed in the MS spectra from RBC samples of patients with thalassaemia major (Jiang, Ma et al. 1994). However, a ferritin-like component is possible to appear in an MS spectrum of a blood sample from a β-thalassaemic mouse, since a thalassaemic mouse does not undergo medical treatment as the thalassaemic patients do. Due to the low S/N ratio in the corresponding spectrum from the thalassaemic mouse such a ferritin-like component was not currently observed, but it will be possible in future samples when the lab-mice diet is enriched with $^{57}$Fe.

**Liver and Spleen**

Figures 4a and 4b shows the MS spectra at 78K of the liver samples from the normal and the thalassaemic mouse, respectively. The spectrum of the thalassaemic sample (Fig. 4b) with an (S/N) of 16 was fitted with a single doublet ($\chi_\nu^2 = 0.99$) while the spectrum of the normal sample (Fig. 4b) with an (S/N) of 5 was not fitted. The MS spectrum of the normal sample presents lower absorption than the thalassaemic sample due to the lower iron concentration. Note that the thalassaemic mouse did not receive any transfusion or any iron-chelation treatment.

Table 3 shows the Mössbauer fitting parameters for the thalassaemic sample. They are in agreement with those for ferritin-like characteristics as obtained, at the same temperature, from dietary-iron-loaded and parenteral-loaded liver samples of rats (Chua-anusorn et al. 1999). In the same study (Chua-anusorn et al. 1999), a small sextet component was observed in the spectra of almost half the liver samples. Even though the sextet signal-to-noise ratio was low, its parameters were in fair agreement with the corresponding ones obtained from a human thalassaemic liver sample (Pierre et al. 1998). According to Bell (Bell et al. 1984), at 80K, the haemosiderin magnetic transition can be observed. Rimbert (Rimbert et al. 1985) observed such a sextet component at 80K only in the spectra of haemodesiderosis liver samples from regularly transfusioned thalassaemic patients but not in the spectra of normal human liver samples or of iron-overload rat samples from an excessive intestinal iron absorption. We also measured a spectra in the velocity range of ±8mm/s but we could not observe such a sextet component. This may be due to smaller iron core in mice in comparison with those in the abovementioned literature.

Figures 5a and 5b shows the MS spectra at 78K of the spleen samples from the normal and thalassaemic mice, respectively. The spectrum of the normal sample (Fig. 5a, S/N=27) was fitted with a single symmetrical doublet ($\chi_\nu^2 = 1.04$). Its Mössbauer parameters, given in Table 3, are in agreement with those of a ferritin component observed in the corresponding MS spectra, at the same temperature, of human spleen samples (Chua-anusorn et al. 1994). Also, Chua-anusorn (Chua-anusorn et al. 1994) did not observe any sextet component in any of the 24 normal human samples.

The corresponding spectrum of the spleen sample for the thalassaemic mouse (Fig. 5b, S/N=39) was fitted with two sub-doublets ($\chi_\nu^2 = 1.01$): a large component (97%) with a slightly asymmetric doublet of a ferritin-like component and another small component (3%) of deoxy-haemoglobin, which is probably due to a remained amount of blood within the spleen. Table 3 shows the Mössbauer parameters for the fit. The extracted values are in agreement with those reported from dietary-iron-loaded and parenteral-loaded spleen samples of rats (Chua-anusorn et al. 1999) and also from thalassaemic human spleen samples (Chua-anusorn et al. 1994). A sextet component was observed in Pierre (Pierre et al. 1998) but not in Chua-anusorn (Chua-anusorn et al. 1999). We also did not observe such a sextet component after measuring our spectrum in the velocity range of ±8 mm/s. This might be due to smaller iron core in mice and/or due to a lower blocking temperature for this sextet component. Both spectra of the thalassaemic liver and spleen samples show an increased ferritin-like iron relatively to the normal ones and this is because thalassaemic tissues exhibit an increased iron deposition due to iron-overload.

**Conclusions**

Mössbauer spectra of blood, liver and spleen samples from a normal and a thalassaemic lab-mouse were acquired at 78K.

The MS spectrum of the normal blood sample was well-fitted with two sub-doublets of the same area, representing the α- and β-subunits of oxy-haemoglobin. The decreased absorption effect observed in the spectrum of the thalassaemic sample is due to the reduced haemoglobin in the sample which indicates an extreme anaemia, as it is expected from a thalassaemia major patient without any treatment (Eleftheriou 2003).

An increased amount of ferritin-like iron was observed in the thalassaemic liver and spleen samples compared to the normal ones as expected since thalassaemic tissues exhibit an increased iron deposition due to the iron-overload. However, the absorption effect from the samples should be treated with some caution since the sample population was consisted from only one mouse of each group.

The Mössbauer fitting parameters obtained from this work are in good agreement with those existing in the literature. As obtained from our results, normal and thalassaemic mice exhibit similar iron complexes to humans, both in their blood and organs. In this sense, our thalassaemic mouse model represents a promising candidate to study thalassaemia with MS.

Future studies will include the enrichment of mice with $^{57}$Fe through their diet to increase the absorption in the MS spectra, thus enabling the characterization of iron complexes. $^{57}$Fe enriched blood and organ samples of normal and thalassaemic mice can be compared at various ages to investigate the rate of iron accumulation.

A direct comparison of normal and thalassaemic mice samples may reveal iron complexes due to thalassaemia which are not affected by blood transfusion and iron-chelators. This will provide a deeper insight into the iron complexes associated with thalassaemia, which may also add to the design of more appropriate iron binding agents.

**Acknowledgments**

We gratefully acknowledge the financial support from the University of Cyprus. The Mössbauer Spectroscopy facility was purchased by national resources, started from the year 2011 (Tender No. PHYS032/11). We also acknowledge the University internal grant 2016 "Postdoctoral Researchers". We would like to thank Prof. Thomas Bakas and Prof. Alexios Douvalis from the University of Ioannina in Greece for their helpful advice in Mössbauer spectroscopy.

**Ethical Approval:** "All applicable international, national and/or institutional guidelines for the care and use of animals were followed."

**TABLES:**

**Table 1.** The Mössbauer fitting parameters using an α-Fe foil (RT) and a Fe powder at 298K and 78K: peak position X(i) and the corresponding line width Γ(i), ), where i is the line number, as well as corresponding calculated X(i) values (Maddock 1998). The statistical errors for X(i) and Γ(i) are ±0.002 and ±0.004 mm/s, respectively. The instrumental error is <0.03mm/s

| Sample | Line number (i) | 1 | 2 | 3 | 4 | 5 | 6 |
|---|---|---|---|---|---|---|---|
| Reference Values (RT) | X(i) [mm/s] | -5.328 | -3.083 | -0.839 | 0.839 | 3.083 | 5.328 |
| α-Fe foil (RT) | X(i) [mm/s] | - | -3.082 | -0.840 | 0.840 | 3.082 | - |
|  | Γ(i) [mm/s] | - | 0.281 | 0.271 | 0.279 | 0.281 | - |
| Fe powder (RT) | X(i) [mm/s] | - | -3.081 | -0.838 | 0.838 | 3.080 | - |
|  | Γ(i) [mm/s] | - | 0.275 | 0.263 | 0.265 | 0.275 | - |
| Fe Powder (78K) | X(i) [mm/s] | - | -3.045 | -0.751 | 0.973 | 3.265 | - |
|  | Γ(i) [mm/s] | - | 0.269 | 0.263 | 0.253 | 0.278 | - |

**Table 2.** The measured FWHM, the net count rate (Rate) and the transmission (T) without a holder and with a Plexiglas (PMMA) and a Polypropylene (PP) holder at $^{57}$Co/Rh source energies of 6.36 and 14.41 keV, in figures 2a-2c.

| | $E_\gamma$=6.36 keV | | | $E_\gamma$=14.41 keV | | |
|---|---|---|---|---|---|---|
| Material | FWHM | Rate [cps] | Trans % | FWHM | Rate [cps] | T% |
| no holder | 0.155 | 124 | 100 | 0.20 | 209 | 100 |
| PMMA holder | - | - | - | 0.22 | 117 | 56 |
| PP holder | 0.153 | 96 | 77 | 0.21 | 201 | 96 |

**Table 3.** The line width (Γ), isomer shift (δ), quadrupole splitting ($\Delta E_Q$), and the relative area of the sub-doubles A and B, if any, from the Mössbauer fitting parameters of figures 3a, 4b, 5a and 5b. The instrumental error is <0.03 mm/sec. The statistical error is also shown.

| Sample | Sub-spectra | Γ [mm/s] | δ [mm/s] | $\Delta E_Q$ [mm/s] | Area [%] |
|---|---|---|---|---|---|
| Normal Blood | A | 0.24±0.03 | 0.28±0.01 | 2.23±0.02 | 51 |
|  | B | 0.40±0.03 | 0.28±0.01 | 1.90±0.07 | 49 |
| Thalas. liver |  | 0.63±0.02 | 0.46±0.01 | 0.71±0.01 | 100 |
| Normal spleen |  | 0.58±0.01 | 0.47±0.01 | 0.69±0.01 | 100 |
| Thalas. spleen | A: Ferritin-like | 0.56±0.01 | 0.47±0.01 | 0.69±0.01 | 97 |
|  | B: deoxy-Hb | 0.24±0.06 | 0.86±0.02 | 2.27±0.03 | 3 |

**FIGURE CAPTIONS:**

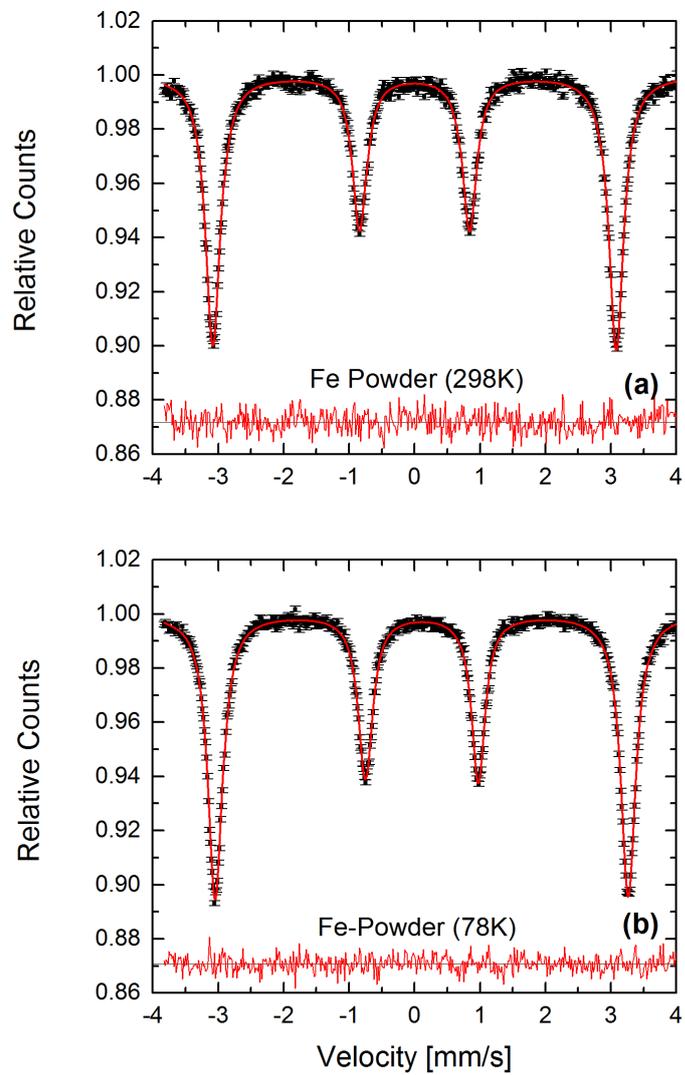

**Fig. 1** The measured Mössbauer spectra of the Fe-powder samples at: (a) RT (298K) and (b) 78K

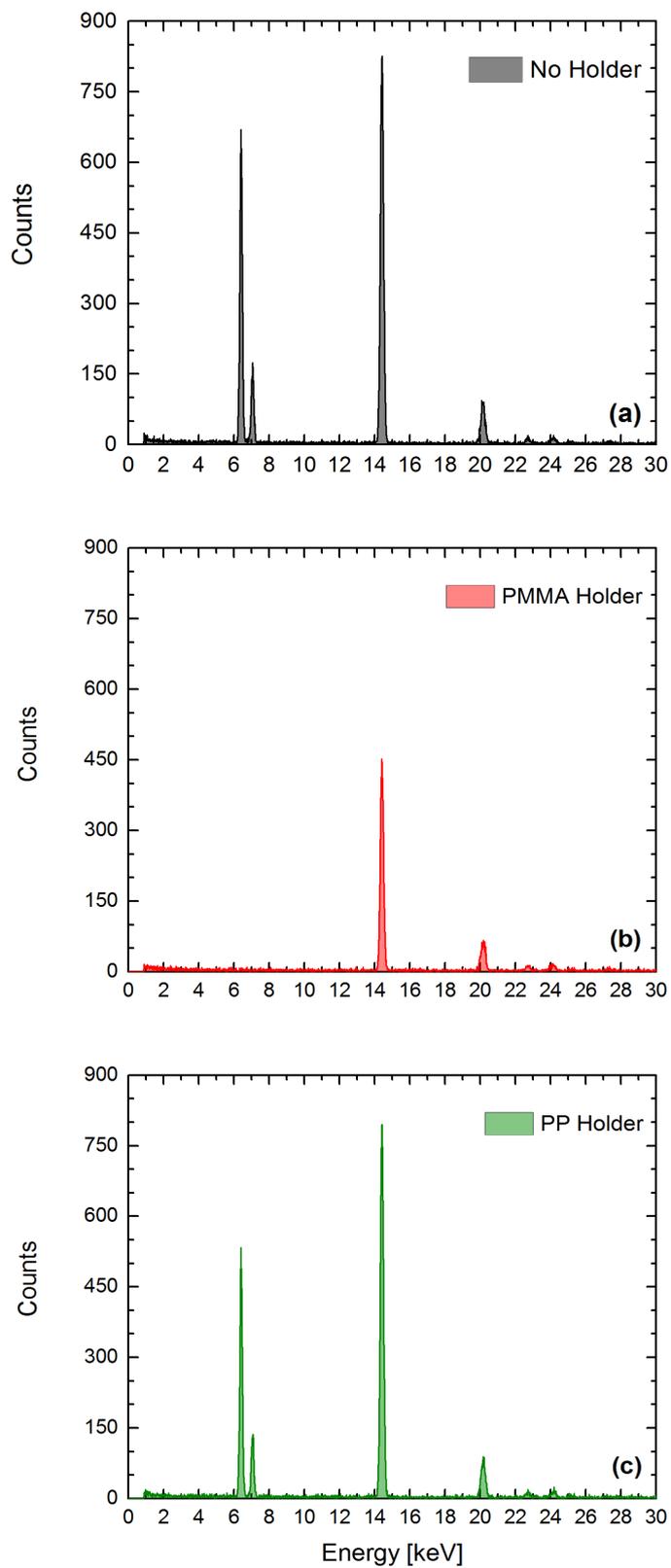

**Fig. 2** The transmission spectra (a) without holder, (b) for a Plexiglas (PMMA) holder, and (c) for a polypropylene (PP) holder for the $^{57}$Co/Rh source energies using a Si-PIN high resolution X-Ray detector.

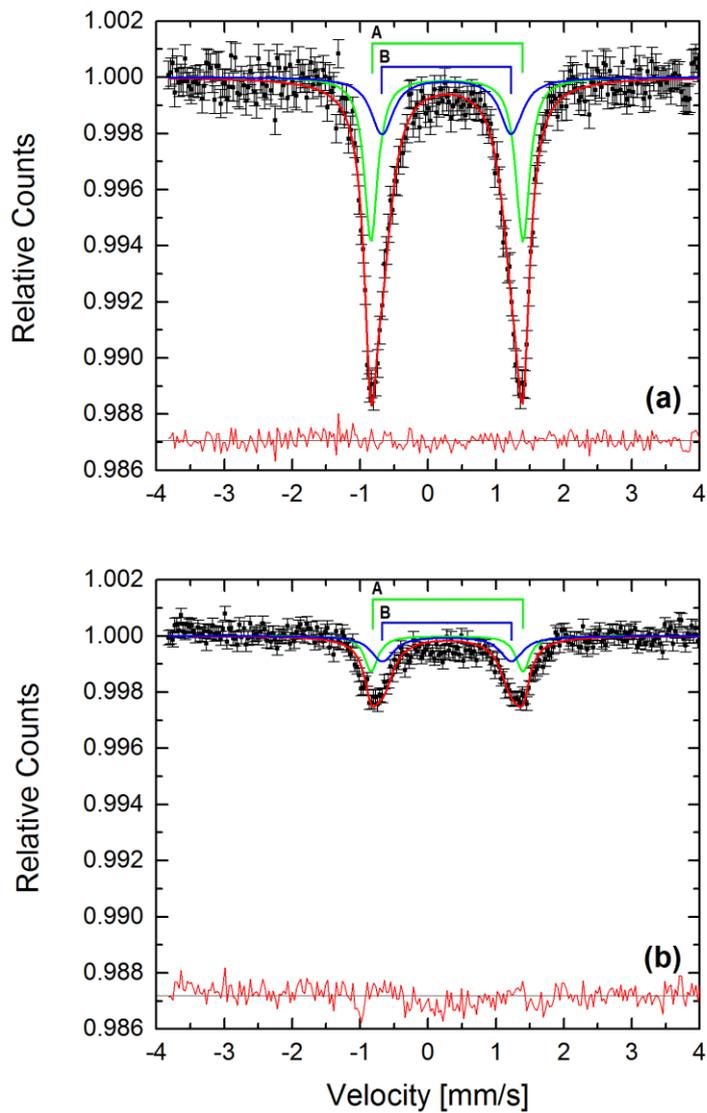

**Fig. 3** The measured Mössbauer spectra, at 78K of blood samples from: (a) one normal and (b) one thalassaemic mice, plotted at the same scale. The α- and β-subunits of oxy-haemoglobin in the fitted spectra are denoted by A and B, respectively. The residuals of the fits are also shown.

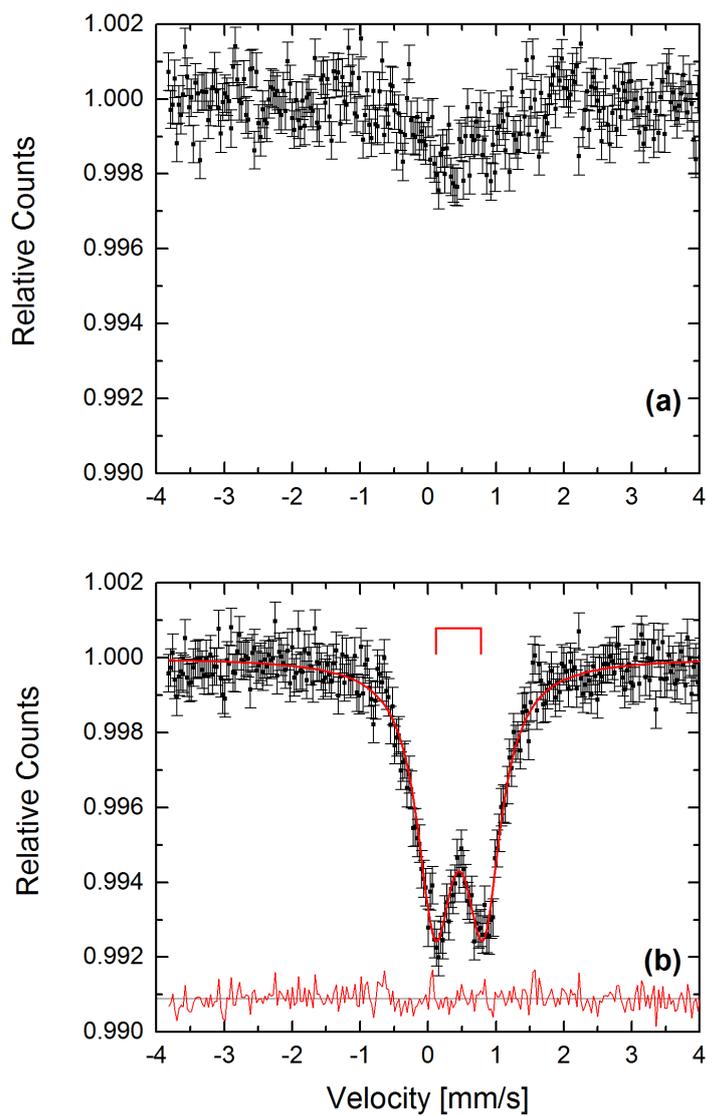

**Fig. 4** The measured Mössbauer spectra, at 78K, of liver samples from: (a) one normal and (b) one thalassaemic mice, plotted at the same scale. The thalassaemic sample is fitted with a doublet. The residuals of the fit are also shown.

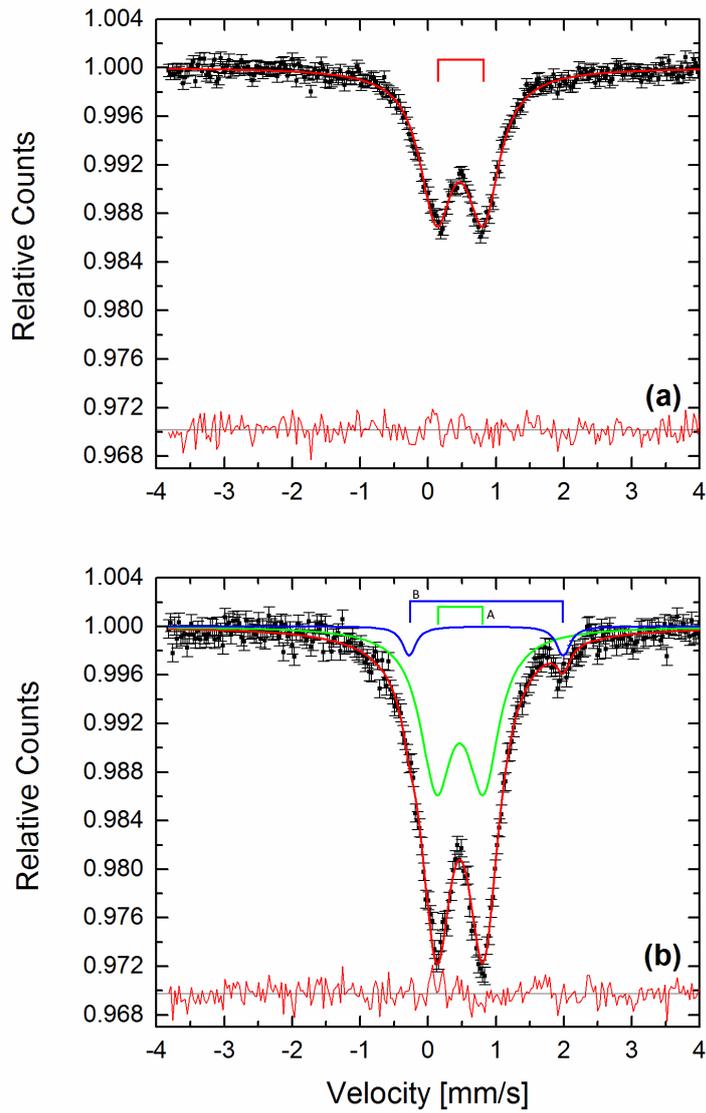

**Fig. 5** The measured Mössbauer spectra, at 78K, of spleen samples from: (a) one normal and (b) one thalassaemic mice, plotted at the same scale. The normal sample is fitted with a single doublet while the thalassaemic sample with two doublets (a small and a large component). The residuals of the fits are also shown.